\documentclass[authorversion,sigconf,10pt]{acmart}

\setcopyright{rightsretained}

\acmDOI{10.1145/3563766.3564089}

\acmISBN{978-1-4503-9899-2/22/11}

\acmConference[HotNets '22]{The 21st ACM Workshop on Hot Topics in Networks}{November 14--15, 2022}{Austin, TX, USA}
\acmBooktitle{The 21st ACM Workshop on Hot Topics in Networks (HotNets '22), November 14--15, 2022, Austin, TX, USA}
\acmPrice{}
\acmYear{2022}
\copyrightyear{2022}

\usepackage{adjustbox}
\usepackage[notcomma,notperiod,notquote,notexcl,notcolon,notscolon]{hanging}
\usepackage{booktabs}
\usepackage[font={small}]{caption}
\usepackage{pifont}
\usepackage{color}
\usepackage{colortbl}
\usepackage{enumitem}
\usepackage{fontawesome5}
\usepackage{framed}
\PassOptionsToPackage{hyphens}{url}
\PassOptionsToPackage{pdfstartview=FitH,hidelinks,pdfpagelabels,bookmarksdepth=3,bookmarksopen=true,bookmarksnumbered}{hyperref}
\usepackage{nicefrac}
\usepackage{mathtools}
\usepackage{microtype}
\usepackage{multicol}
\usepackage{multirow}
\usepackage{scalefnt}
\usepackage{setspace}
\usepackage{thm-restate}
\usepackage{tikz}
\usetikzlibrary{decorations.pathreplacing}
\usetikzlibrary{calc}
\usetikzlibrary{hobby}
\usetikzlibrary{decorations.markings, decorations.pathmorphing}
\usetikzlibrary{shapes}
\usetikzlibrary{positioning,fit}
\usepackage{xcolor}
\usepackage{xspace}
\usepackage{wrapfig}
\usepackage{siunitx}
\usepackage[capitalise,noabbrev]{cleveref}
\usepackage{algpseudocode}
\usepackage[compact]{titlesec}

\newcounter{ExperimentCount}

\newcounter{PropertyCount}

\newif\iffull
\fulltrue

\newif\ifdraftmode
\draftmodefalse 
\ifdraftmode
    \long\def\emma#1{{\color{blue}{\bf Emma: }{\small [#1]}}}
    \long\def\vivian#1{{\color{magenta}{\bf Vivian: }{\small [#1]}}}
    \long\def\raluca#1{{\color{red}{\bf Raluca: }{\small [#1]}}}
    \long\def\natacha#1{{\color{violet}{\bf Natacha: }{\small [#1]}}}
\else
    \long\def\emma#1{}
    \long\def\vivian#1{}
    \long\def\raluca#1{}
    \long\def\natacha#1{}
\fi

\makeatletter
\g@addto@macro{\UrlBreaks}{\UrlOrds}
\makeatother

\newlength{\defhangindent}
\definecolor{LightCyan}{rgb}{0.88,1,1}

\long\def\com#1{}
\long\def\eat#1{}

\long\def\xxx#1{}

\renewcommand{\paragraph}[1]{\smallskip\noindent\textbf{#1}}

\makeatletter
\renewcommand*{\thetable}{\arabic{table}}
\renewcommand*{\thefigure}{\arabic{figure}}
\let\c@table\c@figure
\makeatother 

\setlist[description]{leftmargin=\parindent,topsep=0ex,itemsep=0ex,partopsep=1ex,parsep=1ex}

\LetLtxMacro{\oldtextsc}{\textsc}
\renewcommand{\textsc}[1]{\oldtextsc{\scalefont{1.2}#1}}

\setitemize{noitemsep,topsep=2pt,parsep=2pt,partopsep=2pt,leftmargin=4ex}
\setenumerate{noitemsep,topsep=2pt,parsep=2pt,partopsep=2pt,leftmargin=4ex}

\newcolumntype{R}[2]{%
  >{\adjustbox{angle=#1,lap=\width-(#2)}\bgroup}%
    c%
    <{\egroup}%
}

\crefformat{section}{§#2#1#3}

\begin{document}

\title{Reflections on trusting distributed trust}
\author{Emma Dauterman}
\affiliation{
\institution{UC Berkeley}
\country{}
}
\authornote{Equal contribution.}
\author{Vivian Fang}
\affiliation{
\institution{UC Berkeley}
\country{}
}
\authornotemark[1]
\author{Natacha Crooks}
\affiliation{
\institution{UC Berkeley}
\country{}
}
\author{Raluca Ada Popa}
\affiliation{
\institution{UC Berkeley}
\country{}
}
\renewcommand{\shortauthors}{E. Dauterman, V. Fang, N. Crooks, and R. A. Popa}
\date{}

\frenchspacing
\begin{abstract}
Many systems today distribute trust across multiple parties such that the system provides certain security properties if a subset of the parties are honest.
In the past few years, we have seen an explosion of academic and industrial cryptographic systems built on distributed trust, including secure multi-party computation applications (e.g., private analytics, secure learning, and private key recovery) and blockchains.
These systems have great potential for improving security and privacy, but face a significant hurdle on the path to deployment. We initiate study of the following problem: a single organization is, by definition, a single party, and so how can a single organization build a distributed-trust system where corruptions are independent? 
We instead consider an alternative formulation of the problem: rather than ensuring that a distributed-trust system is set up correctly by design, what if instead, users can audit a distributed-trust deployment?
We propose a framework that enables a developer to efficiently and cheaply set up any distributed-trust system in a publicly auditable way. To do this, we identify two application-independent building blocks that we can use to bootstrap arbitrary distributed-trust applications: secure hardware and an append-only log.
We show how to leverage existing implementations of these building blocks to deploy distributed-trust systems, and we give recommendations for infrastructure changes that would make it easier to deploy distributed-trust systems in the future.
\end{abstract}

%
%



\maketitle
\section{Introduction}

Distributing trust is a powerful tool for building efficient systems with strong privacy and integrity properties.
A distributed-trust system is deployed across $n$ parties (we will subsequently refer to these as trust domains) where, if there are no more than $f$ independent corruptions, the system provides certain security, privacy, and/or integrity properties.
In the past few years, we've seen an explosion of academic and industrial cryptographic systems built on distributed trust; some applications are based on secure multi-party computation~\cite{BGW88,GMW87,Yao82} (e.g. private search~\cite{waldo,dory,senate,splinter}, private analytics~\cite{prio,poplar,exposure-notifications,firefox-telemetry,prio-services} private media delivery~\cite{popcorn}, private blocklist lookups~\cite{checklist}, private DNS~\cite{ODoH}, anonymous messaging~\cite{riposte,express,Chaum88,riffle,atom,dissent}, and cryptocurrency wallets~\cite{unbound,gemini,paxos,sepior,knox,riddle-and-code,fireblocks}), while others are based on Byzantine fault-tolerant consensus and blockchains~\cite{libra,hyperledger,PBFT,hotstuff,hbft,next-700,sbft,zyzzyva,flexible-bft,Lamport11}.

In this paper, we initiate the academic study of an often overlooked challenge that distributed-trust systems face on the path to deployment: \emph{bootstrapping a distributed-trust system is surprisingly difficult.} Distributed-trust systems only provide strong security guarantees insofar as there is no central point of attack that allows an attacker to compromise more than an application-specific threshold of parties.
Many academic works simply assume the existence of multiple non-colluding servers~\cite{dory,waldo,popcorn,splinter,riposte,express,checklist,popcorn}, but an application developer that wants to deploy a distributed-trust system faces difficult questions:
who should have administrative control over the different trust domains, and how do you convince another party that you do not control to run your system?
We study the difficulties developers have faced when deploying distributed-trust systems in \cref{sec:survey}. 

\begin{figure}[t]
    \centering
    \includegraphics[width=\columnwidth]{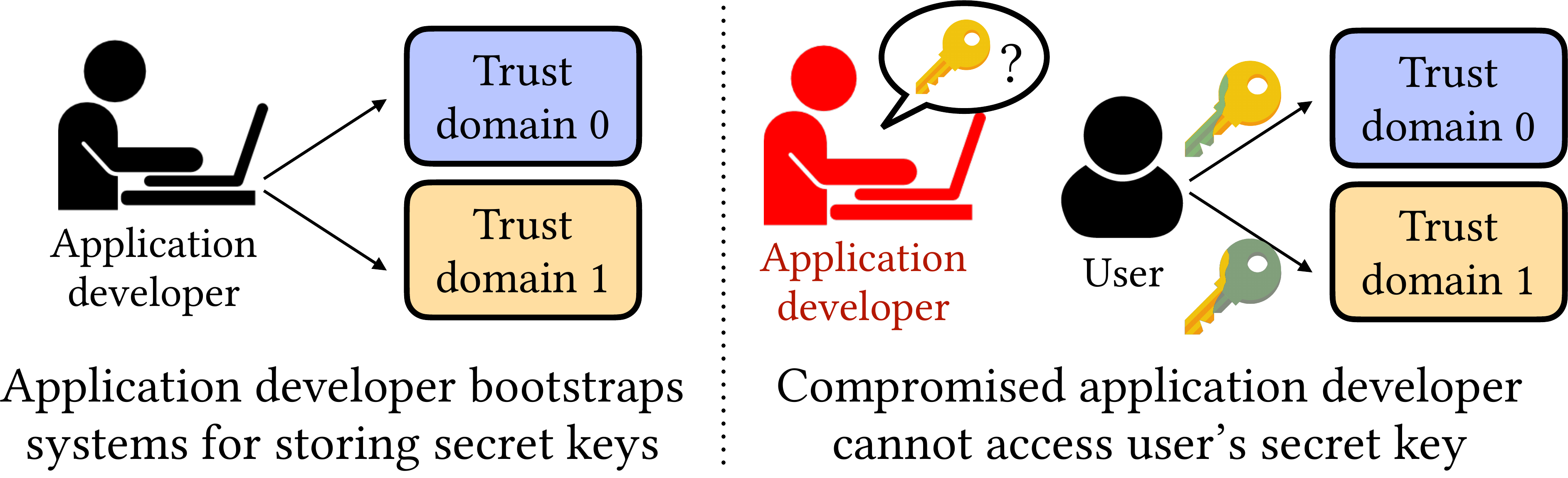}
    \caption{The application developer sets up a distributed-trust system for secret-key backups. The developer should not be a central point of attack in the system.}
    \vspace{-1em}
    \label{fig:intro-fig}
\end{figure}

To make this challenge more concrete, consider a simple application that provides backups for secret keys (e.g., for end-to-end encrypted messaging or cryptocurrency wallets). The user splits its secret key across different trust domains via secret sharing~\cite{Shamir79}. Therefore, even if the attacker steals secret shares from all but one of the trust domains, the attacker cannot learn users' secret keys (\cref{fig:intro-fig}).
As a strawman, the application developer could deploy virtual machines on different cloud providers so that different trust domains correspond to different cloud providers.
The problem with this strawman solution is that the application developer still has administrative control over the virtual machines and so is a central point of attack: if the attacker compromises the developer's credentials, the attacker can easily recover every user's secret key.

To achieve strong security guarantees in practice, we would like trust domains to be truly independent so that there is no single point of failure. 
For security, compromising any system component should always ideally compromise at most one trust domain.
For convenience, a developer should ideally be able to set up a distributed-trust deployment cheaply and easily.
In reality, absolute separation is very hard to achieve given the architecture of today's systems, and so we must settle for some approximation of independence. One good approximation of independence would be to have the application developer coordinate with and cede administrative control to different organizations that manage servers running on different cloud providers with different hardware, potentially even in different geographic locations~\cite{divvi-up-announce, ODoH-cloudflare, firefox-telemetry, enpa-rwc}.
In this way, even if the developer's credentials are compromised, the attacker cannot subvert every trust domain and compromise the entire system.
While this first attempt provides security, it does not provide convenience, as it requires  time-consuming and expensive human coordination across organizations.

Thus we reach an impasse:
\emph{how can an application developer bootstrap a distributed-trust system without herself becoming a central point of attack?}

It is impossible to bootstrap trust out of nothing.
We consider an alternative formulation of the bootstrapping problem: instead of ensuring that a system distributes trust correctly by design, what if users can \emph{audit} a distributed-trust deployment?
We are inspired by the widely deployed certificate transparency infrastructure~\cite{ct} where, instead of preventing certificate authorities (CAs) from issuing bad certificates, users can simply detect CA misbehavior. In our setting, we also provide transparency: rather than guaranteeing that the system always distributes trust correctly, we instead only guarantee that the user will be able to \emph{detect whenever the system does not execute the expected code in different trust domains}.
Moreover, the user will obtain a publicly verifiable proof of misbehavior.

We propose a framework that enables a non-expert application developer to deploy a distributed-trust system in a publicly auditable way without human-level cross-organization coordination.
To do this, we identify two core, application-independent building blocks from which an application developer can bootstrap any distributed-trust application: secure hardware and an append-only log.

To provide public auditability, we use secure hardware to attest to the code running in each trust domain.
Secure hardware such as 
trusted execution environments (TEEs)~\cite{keystone, sanctum, MI6, sgx, nitro} allow the client to verify code integrity (i.e. the secure hardware is running the code that the client thinks it should be running).
To see the history of the code running at each trust domain, the client can check an append-only log maintained by the TEEs.
The developer must publish her code to allow clients and third-party auditors to inspect it and check that it hashes to the value provided by the TEEs.

Because we use secure hardware, it might seem like distributed trust is unnecessary now: we can simply run the entire application inside of secure hardware. 
However, this would make a single type of secure hardware a central point of attack.
For example, if we run the application inside Intel SGX and an attacker finds an exploit in SGX, then they can compromise the security of the entire system (this seems plausible given the history of attacks on SGX~\cite{foreshadow,LVI,RIDL,zombieload,sgxpectre,crosstalk,cacheout,rowhammer-sgx}). Many systems use distributed trust precisely because they do not want security of the entire system to reduce to the security of a TEE. 
To address this issue, we set up our system to split trust across multiple trust domains and use heterogeneous secure hardware.

We demonstrate how organizations can easily bootstrap a distributed-trust application by building on cloud offerings for secure hardware and deployed certificate transparency infrastructure~\cite{ct}.
While it is possible to use only existing infrastructure, we also describe infrastructure-level changes that would make the distributed-trust ecosystem more efficient and sustainable.
To demonstrate the feasibility of our framework, we implement and evaluate a prototype (\cref{sec:eval}).

\paragraph{Limitations.}
Our proposal still has several limitations. The first main drawback is the reliance on secure hardware. We can prevent secure hardware from becoming a central point of attack by using heterogeneous secure hardware.
The second drawback has to do with performance: running an application inside a TEE is more expensive than running it natively (\cref{sec:eval}).
In \cref{sec:design:tomorrow}, we describe how secure hardware manufacturers could design TEEs and how cloud providers could offer services specifically tailored to distributed-trust systems to minimize this overhead.

\section{Distributed-trust deployments today}
\label{sec:survey}

We start by examining the challenges organizations face in deploying distributed-trust applications today. We base our discussion of organizations' solutions on published documentation, including whitepapers and blog posts.

\paragraph{Privacy-preserving analytics.}
Prio~\cite{prio} splits trust across two servers to compute aggregate statistics without revealing individual users' data and has been deployed for Firefox telemetry~\cite{firefox-telemetry} and COVID-19 exposure notification analytics~\cite{exposure-notifications}. For Firefox telemetry, Firefox runs one server and the ISRG (the public-benefit corporation behind Let's Encrypt, which offers free TLS certificates)
runs the other. The COVID-19 exposure notification system computes statistics across iOS and Android users where the ISRG and the National Institute of Health each run a server.
To enable other organizations to easily run a Prio system, the ISRG has announced Divvi Up, a service where the ISRG acts as the second trust domain for a Prio deployment (the organization building the application acts as the ``first trust domain'')~\cite{prio-services,divvi-up-announce}.

While Divvi Up will make it easier for organizations to deploy private analytics systems~\cite{prio,poplar}, it still does not enable general-purpose distributed-trust systems.
Moreover, the challenges ISRG faced in setting up Divvi Up illustrate just how hard it is to set up a distributed-trust system correctly~\cite{divvi-up-lessons,enpa-rwc}. 
For example, debugging and running integration tests now must take place across organizations that don't have a common release process or deployment system.

\paragraph{Digital advertising.}
Meta recently announced their Private Lift Measurement solution~\cite{private-lift} in which Meta and an advertiser run a multi-party computation protocol~\cite{BGW88,GMW87,Yao82}.
Multi-party computation is a cryptographic tool that allows multiple parties to jointly compute some function over their secret inputs such that the parties only learn the output of the computation (and not the other parties' secret inputs). In Meta's use-case, the advertiser can learn how their campaign is doing without revealing unnecessary information to Meta or the advertiser. Even with Meta's resources, Meta reported that it initially took months to onboard a new advertiser~\cite{meta-rwc}.

\paragraph{Private DNS.}
Cloudflare, Apple, and Fastly authored an IETF draft for oblivious DNS over HTTPS that splits trust between a proxy and a resolver such that neither learns both the user's IP address and query~\cite{ODoH-ietf,ODoH-cloudflare,ODoH}. Internet service providers PCCW, SURF, and Equinix have committed to launching proxies,
enabling the set of organizations running proxies to be disjoint from the set of those operating resolvers.

\paragraph{Permissioned ledgers with infrastructure providers.}
To run a new ledger, organizations need to start many nodes quickly. Infrastructure providers like Blockdaemon~\cite{blockdaemon},\\  Alchemy~\cite{alchemy}, or Figment~\cite{figment} offer nodes as a service. For example, Blockdaemon provides end-to-end nodes for permissioned ledgers like Diem~\cite{libra} and Hyperledger Fabric~\cite{hyperledger} (among many others). However, these infrastructure providers are themselves centralized; compromising a provider like Blockdaemon could enable an attacker to compromise a significant fraction of system nodes. The existence of these infrastructure providers illustrates that organizations need a way to easily add nodes to ledger systems.

\paragraph{Financial custody.}
Users transfer cryptocurrency by signing transactions, and so transaction signing keys can secure millions of dollars. Many financial custody companies deploy solutions where the signing key is split across hardware security modules (HSMs), and the HSMs run a multi-party computation to generate a signature on a transaction~\cite{gemini,paxos,sepior,unbound,knox,riddle-and-code,fireblocks}. In this way, no HSM ever holds the entire signing key.

A limitation of the financial custody companies we surveyed is that they only provide security if the company is honest at setup time.
One company deploys and maintains all of the secure hardware, and the end-user cannot check that the system is set up and distributes trust in the way that the company claims.
Furthermore, if the company locks itself out of its machines to defend against post-setup compromise, there is no way to patch bugs or push updates. 

\subsection{Our findings}
The applications we survey fall into one of two categories:
\begin{itemize}
    \item The application setup was challenging, application-specific, and required cross-organization coordination.
    \item The application architecture compromises on distributed trust or functionality in some way (beyond what we could reasonably hope for). 
\end{itemize}
In the first category, we have the ISRG's Divvi Up, Meta's Private Lift Measurement, and oblivious DNS. In all of these cases, the system splits trust across organizations correctly, but the developers had to overcome significant hurdles to coordinate across organizations.
In the second category, we have some permissioned ledger deployments and financial custody solutions. The challenges we see in these deployments illustrate how difficult it is to truly eliminate a central point of attack, especially as the number of parties grows (e.g. for permissioned ledgers).

\paragraph{Going forward.}
While applications in the first category provide strong security guarantees, setting up this type of deployment is simply too difficult for many small organizations. 
In the remainder of this paper, we describe how a developer can set up a distributed-trust application \emph{without expensive, cross-organization coordination}.
\section{System overview}
\label{sec:overview}

We start by describing the building blocks necessary to bootstrap arbitrary distributed-trust applications (\cref{sec:overview:bb}), then describe our system architecture (\cref{sec:overview:arch}), and finally state the properties our system does and does not provide (\cref{sec:overview:properties}).

\subsection{Building blocks}
\label{sec:overview:bb}

Our system requires two core, application-independent building blocks: secure hardware and an append-only log.

\paragraph{Secure hardware.}
Secure hardware should be able to attest to the code that is running.
In particular, the client should be able to verify that it is communicating with a correctly provisioned piece of secure hardware running software that hashes to a particular value. 
In addition, if a developer is running an integrity-preserving application, the secure hardware should isolate memory or detect tampering, and for privacy-preserving applications, the secure hardware should also encrypt the memory contents.
Existing industrial~\cite{sgx,nitro} and academic~\cite{sanctum,keystone,MI6} TEEs provide these properties (note that because TEEs generally do not protect memory access patterns~\cite{BMDK+17,HCP17,cachezoom,SWG+17,branch-shadowing,no-page-faults,controlled-channel,membuster}, developers writing privacy-preserving applications should ensure that memory access patterns do not reveal secret data).
In \cref{sec:design:tomorrow}, we discuss future alternatives to TEEs.

\paragraph{Append-only log.}
The append-only log should provide integrity: once an entry is added, it cannot be altered or deleted.
Distributed-trust applications require at least one honest trust domain to provide meaningful security guarantees, and so we can build an append-only log by having each TEE maintain a copy of this log.
To query the log, the client can simply query each TEE and check that the responses match.

\subsection{System architecture}
\label{sec:overview:arch}

\begin{figure}
    \centering
    \includegraphics[width=\columnwidth]{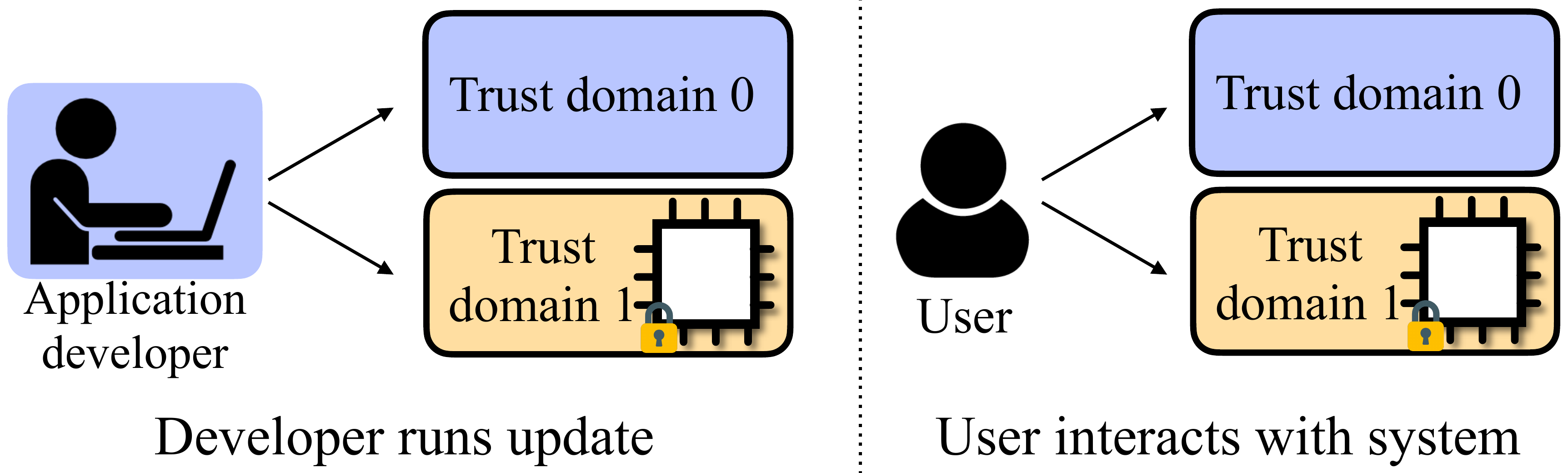}
    \caption{System architecture. We show two trust domains for simplicity, but in practice the number of trust domains is application-dependent. Trust domain 0 is run by the application owner without any secure hardware.}
    \vspace{-1em}
    \label{fig:sys-arch}
\end{figure}

We describe the architecture of a system that distributes trust across $n$ trust domains (\cref{fig:sys-arch}).
Each trust domain runs a server equipped with secure hardware. Ideally trust domains should leverage different types of secure hardware to minimize the chance that an exploit in one type of secure hardware compromises the entire system.
The application developer can run one trust domain on her own without any secure hardware (denoted as ``trust domain 0'' in \cref{fig:sys-arch}).

Note that for applications with a small number of trust domains, it is possible to run each trust domain on a different type of secure hardware. For applications with a large number of trust domains, the fact that there are a comparatively small number of secure hardware vendors means that some trust domains might have to use the same type of secure hardware, potentially introducing correlated corruptions (the application only provides security or privacy if the attacker can compromise more than an application-specific threshold of trust domains).
We discuss how to improve this shortcoming moving forward in \cref{sec:design:tomorrow}.

\subsection{System properties}
\label{sec:overview:properties}

Given a distributed-trust application with $n$ trust domains that provides security if at least $t$ are honest, we provide the following guarantees.

\paragraph{Auditable.}
For each of the $n$ trust domains, the client can obtain a digest of the code that is currently running and a history of digests corresponding to code that ran previously.
The client can check that the digests match across all $n$ trust domains, ensuring that if at least one trust domain is honest (as required by distributed-trust applications), the client will receive a digest of the correct code.
The developer open-sources her code so that clients and third-party auditors can inspect the published code to make sure that it does what the developer claims and check that the hash of the published code matches the hash from the TEEs. We provide security if the published code is running correctly in at least $t$ trust domains.

\paragraph{Simple for application developer.}
Our system enables application developers to bootstrap distributed-trust systems using existing cloud resources and infrastructure. Human-level cross-organization coordination is not necessary. 

\paragraph{Supports code updates.}
Application developers can securely update their code. Code updates are necessary to fix security-critical bugs and support new features. Clients learn when the code running in different trust domains is updated, and they can check that the new code matches the hash in the log. Moreover, clients can check that the trust domains correctly update to the new code when the developer publishes an update. Because the code is open-source, clients and third-party auditors can check the contents of the current code and old versions of the code.

\paragraph{Non-goals.}
We do not defend against implementation bugs or backdoors. By examining the application code, clients and third-party auditors can gain confidence that the code is doing what the application developer claims, but different implementations or formal verification would be necessary to protect against correlated compromise due to the implementation.
Similarly, we can only provide limited protection in the case where the developer is herself the attacker and is allowed to push code updates; the developer can insert backdoors that could be very challenging to detect. Therefore for highly sensitive applications, a developer might consider disabling her ability to push code updates to defend against future compromise.

\section{System design}
\label{sec:design}

We now describe our system design. We first show how to bootstrap distributed trust today (\cref{sec:design:today}). We then outline infrastructure-level changes that would better enable bootstrapping in the future (\cref{sec:design:tomorrow}).

\subsection{Deployment today}
\label{sec:design:today}

To bootstrap distributed trust today, we can build on top of a TEE like Intel SGX or AWS Nitro. Cloud providers already provide access to TEEs, with Microsoft Azure supporting Intel SGX and AWS supporting the Nitro enclave,
and so using the techniques we describe below, an application developer can deploy a distributed-trust application immediately.
As a starting point, we first explain how deployment without updates works, and then we show how to layer on support for updates. 

\paragraph{Starting point: deployment without updates.}
Without updates, the system design is straightforward: the developer seals the application code directly inside a TEE in each trust domain. The client can then use the TEE's attestation mechanism to receive a hash of the sealed code from each trust domain. If all the hashes match, the client knows that the trust domains all claim to run the same code and, if $t$ trust domains are honest, the hash must correspond to the code currently running in $t$ trust domains. The client (or a third-party auditor) can optionally inspect the open-source code corresponding to the hash; if enough clients or third-party auditors inspect the code, other clients will generally have confidence in the deployment.
Because the code is sealed onto the enclave, the application developer cannot change the code, which provides security, but also makes code updates impossible. 

\paragraph{Supporting updates.} 
Application developers need the ability to fix security-critical bugs and support new features. Supporting updates with current TEEs is challenging because existing TEEs do not support updating the existing code while maintaining the current state of the running application.
Moreover, the client needs some way to learn when an update has happened and check that the update was performed correctly. Because we need to defend against malicious updates, we cannot make any assumptions about the behavior of the new code (e.g., we cannot assume that the new code will correctly alert the client that an update occurred). 

To address these problems, we add a layer of indirection. Instead of sealing the developer's code directly on to the enclave, we instead seal an \emph{application-independent framework} on to the TEE. This application-independent framework accepts application code as input and executes it. When the application developer wants to update the code running, she sends the new code to the TEE. Before the TEE starts running the new code, it alerts the client that an update is about to take place and sends the user a hash of the updated code. We can open-source this application-independent framework in order to increase confidence in this framework, and, as before, the client can use attestation to verify that the framework is running correctly on the TEE.

We need to ensure that a malicious update does not prevent the application-independent framework from notifying the client that an update took place. To ensure that the update cannot interfere with the framework, we run the updated application code inside of a sandbox~\cite{webassembly,nacl}.
Sandboxing the application code ensures that the executed code cannot ``escape'' the sandbox and have an effect on the system outside the sandbox (i.e. the framework). We also need to ensure that the TEE only runs updates from the application developer. We can do this easily by sealing on to the TEE not just the framework, but also a public key. Then each subsequent update needs to be accompanied by a signature that verifies under the original public key.

In order to ensure that a malicious developer cannot erase evidence of malicious code, each TEE maintains an append-only log of code digests. This append-only log (implemented at each TEE as a hash chain) allows clients and third-party auditors to query for and audit old code digests.
Prior work has explored transparency logs for application binaries,
but in the context of verifying local client code rather than remote server code~\cite{contour, chainiac, sw-transparency}.

\subsection{Deployment tomorrow}
\label{sec:design:tomorrow}

Infrastructure changes would make deployment even easier and provide greater flexibility in the future.

\paragraph{Expanding cloud provider offerings.}
Cloud providers should offer services specifically tailored for distributed-trust systems. In particular, cloud providers should allow developers to submit code and code updates and then run the code without allowing the developer to inspect or modify application memory. The cloud provider should attest to the current code that is running, as well as the history of executed code.
In this way, the client could gain some confidence that the correct code is running and that the application developer cannot access application memory without secure hardware.

\paragraph{Secure hardware design.}
TEEs like SGX support sealing code on to the TEE. This design decision requires us to seal our general-purpose framework on to the TEE and then have our framework run the dynamic application code in a software-based sandbox. Changing the hardware design could allow us to support updates much more efficiently.
Instead of running the new binaries inside a software sandbox inside the TEE, the hardware could instead isolate the framework from the application binary directly. We simply need the secure hardware to attest to the framework that is running, store a history of executed code, and provide a mechanism for the framework to effectively sandbox the new binary.
We hope that our work spurs the development of secure hardware explicitly tailored to bootstrapping distributed-trust systems efficiently.

\section{Evaluation}
\label{sec:eval}

We implemented a prototype of our framework and support execution on AWS Nitro.
We use WebAssembly (Wasm)~ \cite{webassembly} as our sandboxed execution environment. 
We compile C++ applications into Wasm using Emscripten~\cite{emscripten} and run Wasm applications inside Node.js~\cite{nodejs}. 
We implement a BLS threshold signature~\cite{bls} application on top of our framework using libBLS~\cite{libbls}: each trust domain stores a secret key share, and the trust domains can jointly sign a message.

We evaluate our framework on a single AWS \texttt{c5.4xlarge} instance with a 16-core Intel Xeon 8124M CPU and 32GB RAM. We allocate 4GB RAM and 2 cores for the Nitro TEE.

\begin{table}[t]
    \centering
    \begin{tabular}{ccc}
        \toprule
            Execution Environment & Processing Time & Increase \\
            \midrule
            Baseline & 10.2ms & --- \\
            Sandbox & 14.9ms & 46.1\%\\
            TEE + Sandbox & 15.8ms & 54.9\% \\
        \bottomrule
    \end{tabular}
    \caption{Processing time for producing a BLS threshold signature share under different execution environments. The baseline corresponds to native execution (no TEE and no sandbox).}
    \vspace{-1em}
    \label{tab:overhead}
\end{table}

\paragraph{Framework overheads.}
Table~\ref{tab:overhead} shows the threshold signing time for different execution environments. Our baseline measures the processing time of the C++ implementation without a TEE and without sandboxing.
Compiling to Wasm and running inside of Node.js imposes a 46.1\% overhead (comparable to a previous study on Wasm performance~\cite{jangda2019not}).
Running the sandboxed application inside the AWS Nitro TEE increases processing time by 54.9\%. This overhead is due to the fact that we need two additional sockets: one to forward request traffic from the client to our framework, and one inside the TEE to communicate between our framework and the sandboxed application.

\section{Conclusion}
Previously, distributed-trust systems were only an option for organizations that could successfully coordinate with other organizations. However, bootstrapping without cross-organization coordination can enable small organizations to securely deploy distributed-trust systems. For example, end-to-end encrypted messaging applications could use distributed trust to establish a public-key infrastructure or back up secret keys (each trust domain stores a secret key share).
We hope that our work motivates the study of building blocks (i.e. secure hardware and append-only logs) tailored specifically to distributed-trust systems.
We also leave to future work the question of if these building blocks are necessary for bootstrapping distributed trust, or if there are a completely different set of building blocks we could leverage instead.
What other trade-offs can developers make to securely bootstrap distributed-trust systems without requiring cross-organization coordination?

\paragraph{Acknowledgments.}
We thank the anonymous reviewers for their helpful feedback. We also thank Miles Wada for participating in early stages of this work, as well as Narek Galystan, Jack Humphries, Aurojit Panda, Samyu Yagati, Wen Zhang and students in the Sky security group for feedback that improved the presentation of the paper. 
This work is supported by NSF CISE Expeditions Award
CCF-1730628, NSF CAREER 1943347, and gifts from the
Alibaba, Amazon
Web Services, Ant Group, Astronomer, Ericsson, Facebook, Futurewei,
Google, IBM, Intel, Lacework, Microsoft, Nexla, Nvidia, Samsung, Scotiabank, Splunk, and
VMware. 
This work is also supported by NSF Graduate Research Fellowships and a Microsoft Ada Lovelace Research Fellowship.


{
\bibliographystyle{plain} 
\bibliography{refs}
}

\end{document}